\documentclass[12pt]{article}

\usepackage{graphicx}

\newcommand{\beq}{\begin{equation}}
\newcommand{\eeq}{\end{equation}} 
\newcommand{\beqa}{\begin{eqnarray}}
\newcommand{\eeqa}{\end{eqnarray}}
\newcommand{\nn}{\nonumber}

\def\ifmath#1{\relax\ifmmode#1\else$#1$\fi}

\def\to{\rightarrow}
\def\from{\leftarrow}

\def\CP{\ifmath{C\!P}}
\def\BR{\mbox{{\rm BR}}}
\def\Re{\mbox{{\rm Re}}}
\def\Im{\mbox{{\rm Im}}}
\def\gsim{{~\raise.15em\hbox{$>$}\kern-.85em
          \lower.35em\hbox{$\sim$}~}}
\def\lsim{{~\raise.15em\hbox{$<$}\kern-.85em
          \lower.35em\hbox{$\sim$}~}}

\begin{document}


\title{\bf $B_d^0(t)\to DPP$ time-dependent Dalitz plots,\\
       \CP-violating  angles $2\beta$, $2\beta +\gamma$,\\
       and discrete ambiguities}
\author{J. Charles,  A. Le Yaouanc, L. Oliver, O. P\`ene and J.-C. Raynal}
\maketitle

\begin{center}
Laboratoire de Physique Th\'eorique et Hautes \'Energies
\footnote{Laboratoire associ\'e au Centre National de la Recherche
Scientifique - URA D00063\\
charles@qcd.th.u-psud.fr, leyaouan@qcd.th.u-psud.fr, oliver@qcd.th.u-psud.fr}
\\
Universit\'e de Paris XI, B\^atiment 211, 91405 Orsay Cedex, France
\end{center}

\thispagestyle{empty}\setcounter{page}{0}

\vskip 5 mm
\begin{flushright}
LPTHE-Orsay 97/70\\
hep-ph/9801363
\end{flushright}

\begin{abstract}
We study \CP-violation in resonant three-body $B_d\to DPP$ decays, where $PP$ stands for either $\bar D\pi$, $\bar DK_S$, $\pi\pi$ or $\pi K_S$. Analogously to the $B_d\to 3\pi$ channel and the extraction of $2\alpha$, the first three channels are shown to measure $\cos 2\beta$ in addition to $\sin 2\beta$, thus allowing to resolve the $\beta\to \frac \pi 2 -\beta$ ambiguity, while the $D\pi K_S$ final state leads to a measurement of $2\beta+\gamma$.
The $B_d^0(t)\to D^+ D^- \pi^0$ channel via the interference between $D^{\ast\ast}$ orbitally excited
 resonances is taken as an example, although this Cabibbo-suppressed decay suffers from irreducible penguin uncertainties. Then two penguin-free and Cabibbo-dominant modes are proposed: $B_d^0(t)\to D^+D^-K_S$ with the $D_s^{\ast\ast}$ resonances, and $B_d^0(t)\to D^0_{\CP} \pi^+\pi^-$ with the $D^{\ast\ast}$ plus the $\rho$.
Finally, the $B_d^0(t)\to D^{\pm} \pi^{\mp} K_S$ channel with the $D_s^{\ast\ast}$ and $K^*$ resonances provides a new clean method to measure the unusual angle $2\beta + \gamma$.
We present in all cases a crude estimate of the number of $\cos 2\beta$ (respectively $2\beta+\gamma$) sensitive
 events. We show that this number is an increasing function of the resonance mass, a favorable situation compared to the more extensively studied three-pion Dalitz plot. However, the poor detection efficiency of
the $D$ mesons could pose a problem. 
As an annex and speculative application of these Dalitz plot based methods, the penguin-dominated $B_d^0(t)\to K_SK_SK_L$ decay also measures $2\beta$.
 \end{abstract}

\newpage

\section*{Introduction}

There is little doubt that the \CP-asymmetry in $B\to J/\psi K_S$
\footnote{Throughout this paper, $B$ means a $B_d$ meson.} will be measured in the next years with an increasing accuracy~\cite{reviewBF}, and will provide 
$\sin 2\beta$ without sizeable hadronic uncertainties. However, measuring only $\sin 2 \beta$ leaves ambiguities, in particular the ambiguity $\beta \to \frac \pi 2 -\beta$. It might be argued that this is not a real problem since 
  the overall constraint on the Unitarity Triangle (UT) already leave no room for this 
  ambiguity, indeed we know that $\cos2\beta>0$~\cite{reviewBF}.
However, if $\beta$ turns out to be not too far from $\pi/4$ an information on the sign of $\cos 2 \beta$ will be important, and furthermore,
 the UT constraints only hold within the Standard Model, which  obviously has
  to be tested with care in $B$-factories.
 
The analogous ambiguity $\alpha \to \frac \pi 2 - \alpha$ was argued in~\cite{quinn} to be
 solvable via the use of the time-dependent $B^0(t)\to \pi^+ \pi^- \pi^0$ Dalitz plot in the $\rho$ resonance region. The interference between, say, the $B^0\to \rho^+\pi^- \to \pi^+ \pi^- \pi^0$ and  the $\bar{B^0} \to \rho^-\pi^+ \to \pi^+ \pi^- \pi^0$ decays and the use of the Breit-Wigner resonant phase allows to measure a term proportional to $\cos 2\alpha$ and thus resolve the ambiguity. The partial understanding of the strong phase in the resonant region and the use of the isospin analysis furthermore provides some hope of separating the penguin contribution (which is far from negligible in this case and does introduce a significant shift on $\alpha$) from the tree one~\cite{quinn,babarbook}.  
 
One may wonder if the same technique may be applied to the measure of the angle $\beta$. 
The straightforward generalization of the $\rho \pi \to 3\pi$ analysis would  be a 
$D^\ast \bar  D\to D^+D^- \pi^0$ one.
However, the interest of the Dalitz plot analysis relies on the interference region,
 say, between the $\rho^+$ and $\rho^-$ bands. This interference region is rather 
 large thanks to the sizeable width of the $\rho$ resonance. The $D^\ast$ is much too narrow ($\Gamma < 0.1$ MeV) to leave any hope. Therefore we considered an analysis
of  $\bar B^0\to D^+D^{\ast\ast -}\to D^+D^-\pi^0$ interfering with $\bar B^0\to D^{\ast\ast +}
 D^-\to D^+D^-\pi^0$ and with the \CP-conjugate channels, where $ D^{\ast\ast}$ represents the
  orbitally excited positive parity 
 charmed mesons ($D_0^\ast$, $D_1^\ast$, $D_1$, $D_2^\ast$). In the final three-body state we can at will replace $D$ 
 by $D^\ast$. In this letter we present a preliminary study of this proposal.
We find that in principle $\cos 2\beta$ can be measured this way, up to irreducible penguin uncertainties 
that cannot be removed via the Dalitz plot, due to the $\Delta I=1/2$
isospin structure of the decay.

We therefore present also two other proposals, free of penguin uncertainty and moreover Cabibbo dominant: $B^0\to D^-D_s^{\ast \ast +} \to D^+D^-K_S$ interfering with
$\bar B^0\to D^+D_s^{\ast \ast -} \to D^+D^-K_S$. This decay is the ``class-I'' analogous
of $B\to J/\psi K_S$. Another interesting channel is $B^0\to D^{\ast\ast -}\pi^+ \to D^0_{\CP}\pi^+\pi^-$ interfering with 
$\bar B^0\to D^{\ast\ast +}\pi^- \to D^0_{\CP}\pi^+\pi^-$ and $B^0(\bar B^0)\to D^0_{\CP}\rho^0 \to D^0_{\CP}\pi^+\pi^-$, where 
$D^0_{\CP}$ means a neutral $D$ decaying 
into a \CP-eigenstate. For example, $D^0(\bar  D^0)\to \pi\pi$, $D^0(\bar  D^0)\to K\bar K$, $D^0(\bar  D^0)\to K_S \pi^0$ etc.

All above mentioned decay channels measure $2\beta$ since, in Wolfenstein
parame\-tri\-zation~\cite{wolf}, the decay amplitudes are \CP-invariant and the $B^0-\bar
B^0$ mixing forwards a $\exp[-2 i \beta]$ phase. The Dalitz plots for $B\to \pi D_s^{\ast\ast} + DK^\ast \to D^{\pm}
\pi^{\mp} K_S$, on the contrary, contain the phase $\gamma$ of $V_{ub}$ and give
 a measure of $2 \beta +\gamma$. Such a measure would be very important, not only as an attempt at the measure of $\gamma$ but also to reduce further the angle ambiguities. Other methods to lift discrete ambiguities in CKM angles were discussed in~\cite{kayser,quinn2}.
 
Unhappily it is not clear if the statistics will allow these analyses to be done in $e^+e^-$ $B$-factories (\textsc{BaBar}, Belle), mainly because of the small detection efficiency of the $D$ mesons in the final states. However, except $B\to D^+D^-\pi^0$, all the presented $B\to DPP$ decays can be detected with charged
particles only, which is a good starting point for hadronic machines (LHC-B, BTeV). 

This letter is organized as follows: in Section~\ref{ddpi} we discuss the case of $B\to D^+D^-\pi^0$ explicitly and define an Effective Branching Ratio, which describes roughly the number of events generated by the Breit-Wigner interference effects and sensitive to the CKM angles. In Sections~\ref{ddk},~\ref{dpipi},~\ref{dkpi} we describe briefly the $B\to D^+D^-K_S$, $B\to D^0_{CP}\pi^+\pi^-$ and $B\to D^{\pm}\pi^{\mp}K_S$ channels. In Section~\ref{kkk} a speculative idea on $B\to K_SK_SK_L$ decay is presented. Finally in Section~\ref{facto} we make some model calculation of the various effects described here.

\section{The $B\to D^+D^-\pi^0$ Dalitz plot} \label{ddpi}

\subsection{The $\cos 2\beta$ dependence}

We will take the example of $B\to D^+ D^- \pi^0$ just to be specific, and we neglect penguins. In Wolfenstein parametrization the weak decay goes through $b \to c\bar cd$, and contains no \CP-odd phase. The only phase in the problem will be the $2 \beta$ phase of 
$B^0 - \bar  B^0$ mixing. Up to trivial angular rotations, two independent variables label the final state, which may usefully be taken as the Dalitz plot variables:
\begin{equation} s^+= (p_{D^+} +p_{\pi^0})^2,\quad s^-= (p_{D^-} +p_{\pi^0})^2,
\quad s^0=(p_{D^+} + p_{D^-})^2, \end{equation}
where the relation
\begin{equation} s^++s^-+s^0=m_B^2 + 2 m_D^2 + m_\pi^2 \end{equation}
tells that there are only two independent variables, say $s^+$ and $s^-$.
\CP-eigenstates are on the line $s^+=s^-$.
Let us define
\begin{equation}
{\cal A}(s^+,s^-) \equiv A(B^0\to D^+ D^- \pi^0),\qquad 
{\bar  {\cal A}}(s^+,s^-) \equiv A(\bar  B^0\to D^+ D^- \pi^0),\label{AAbar}
\end{equation}
where $B^0$ and $\bar  B^0$ represent unmixed neutral $B$.
The amplitudes in~(\ref{AAbar}) contain no weak phase, as already stated,
they do contain unknown strong (\CP-even) phases, which obviously depend on $s^+$ and $s^-$ and are different in ${\cal A}$ and $\bar{{\cal A}}$ except when $s^+=s^-$.

The time-dependent amplitudes for an oscillating state $B^0(t)$ which has been tagged as a $B^0$ meson at time $t=0$ is given by 
\beq
A(s^+,s^-;t) = {\cal A}(s^+,s^-) \cos\left(\frac{\Delta m\,t}2\right) + i e^{-2i \beta}
\bar  {\cal A}(s^+,s^-) \sin\left(\frac{\Delta m\,t}2\right)\,,
\eeq
and the time-dependent amplitude squared is:
\beq
|A(s^+,s^-;t)|^2 =\frac{1}{2}\left[ {\rm G}_0(s^+,s^-)+{\rm G}_{\rm c}(s^+,s^-)\cos\Delta m\,t-{\rm G}_{\rm s}(s^+,s^-)\sin\Delta m\,t \right] \, ,
\label{osc}
\eeq
with
\begin{eqnarray}
{\rm G}_0(s^+,s^-)     & = & |{\cal A}(s^+,s^-)|^2 +|\bar{{\cal A}}(s^+,s^-)|^2 ,\\
{\rm G}_{\rm c}(s^+,s^-) & = & |{\cal A}(s^+,s^-)|^2 -|\bar{{\cal A}}(s^+,s^-)|^2 ,\\
{\rm G}_{\rm s}(s^+,s^-) & = & 2\Im\left (e^{-2i \beta} \bar{\cal A}(s^+,s^-){{\cal A}^\ast(s^+,s^-)} \right ) \nn \\
& = & -2 \sin(2\beta)\, \Re \left ( \bar{\cal A}{{\cal A}^\ast} \right ) + 2\cos (2\beta) \,\Im \left ( \bar{\cal A}{{\cal A}^\ast} \right ).\label{gmix}
\end{eqnarray}
The transformation defining the \CP-conjugate channel $\bar B^0(t)\to D^-D^+\pi^0$ is $s^+\leftrightarrow s^-$, ${\cal A}\leftrightarrow \bar{\cal A}$ and $\beta\to -\beta$. Then:
\beq\label{oscbar}
|\bar A(s^-,s^+;t)|^2 =\frac{1}{2}\left[ {\rm G}_0(s^-,s^+)-{\rm G}_{\rm c}(s^-,s^+)\cos\Delta m\,t+{\rm G}_{\rm s}(s^-,s^+)\sin\Delta m\,t \right]\,.
\eeq
Note that for simplicity the $e^{-\Gamma t}$ and constant phase space factors have
been omitted in Eqs.~(\ref{osc}) and~(\ref{oscbar}).

Thus, if ${\rm G}_{\rm c}(s^+,s^-)$ is not antisymmetric in $(s^+,s^-)$ there is direct \CP-violation (proportional to $\cos\Delta m\,t$); similarly, if ${\rm G}_{\rm s}(s^+,s^-)$ is not antisymmetric there is mixing-induced \CP-violation (proportional to $\sin\Delta m\,t$). Neglecting penguins, no direct $\CP$ occurs, as $\bar{\cal A}(s^+,s^-)={\cal A}(s^-,s^+)$. Furthermore, if the final state was a \CP-eigenstate, ${\rm G}_{\rm s}$ would only contain the mixing-induced \CP-violating $\sin 2 \beta$ term in~(\ref{gmix}). The \CP-conserving $\cos 2 \beta $ term is the new one. Measuring this term resolves the $\beta \to {\pi \over 2} - \beta$ ambiguity \footnote{It does not resolve the $\beta \to \pi + \beta$ ambiguity which requires some theoretical input~\cite{quinn2}.}.  But such a measurement necessitates some knowledge of the \CP-even phase of $\bar{\cal A}(s^+,s^-){\cal A}^\ast(s^+,s^-)$. In general the strong phases are utterly unknown a fortiori in a three-body decay. However the trick, first proposed in the three-pion case~\cite{quinn}, is to assume resonance dominance, at least in chosen parts of the Dalitz plot, and to use the Breit-Wigner phase.

We will consider the orbitally excited $D^{\ast\ast}$ and $\bar  D^{\ast\ast}$ resonances. The obvious problem here is that besides the Breit-Wigner phase, many other sources of strong phases can be thought of, for example due to the final state $D^{\ast\ast}$ interaction with the $\bar  D$, etc. Nobody knows how to compute these phases. What we may however assume is that these background phases vary slowly under the resonances, where the Breit-Wigner phase varies quickly. Thus a fit with a constant background phase plus the Breit-Wigner phases should do the job. The second objection stems from the existence of many other $D\pi$ ($\bar  D\pi$) resonances: higher excited charmed mesons. One cannot use too many parameters in the fit. It is assumed that one can minimize the effect of the unwanted resonances by selecting properly the domain in the Dalitz plot under the
first orbitally excited $D^{\ast\ast}$ and $\bar  D^{\ast\ast}$ resonances.
In our analysis we did not either consider, for simplicity, the charmonia 
$D\bar  D$ resonances. A recent application of these ideas in the non-$\CP$ studies of the $D$ and $D_s$ three-body decays may be found in \cite{frabetti}.

In the following we will use for the masses (in GeV):
\begin{equation}
 m_B=5.279,\quad m_{D}=1.865,\quad
m_{D^{\ast}_0}=2.360,\quad
m_{D_2^\ast}=2.459.
\label{masses}\end{equation}
Since the $J=1$ orbitally excited mesons cannot decay into $D\pi$ (parity plus angular momentum conservation), we restrict ourselves to the $J=0,2$ ones. The $J=2$, $D^\ast_2(2459)$ has been observed and is narrow ($23\pm 5$ MeV) as expected since it decays through a D-wave. In our calculations we have taken $\Gamma_2=20$ MeV. The $D^\ast_0$ 
decays through an S-wave and is expected to be broad, which is enough to explain that it has not yet been observed. We will assume $\Gamma_0=150$ MeV~\cite{kaidalov}.

For the sake of simplicity let us first consider only the contributions 
of the $J=0, D_0^\ast$ and $\bar  D_0^\ast$  resonances
\footnote{It should be stressed that in a real analysis, one would take into account
as many resonances as possible. See, for example, ref.~\cite{frabetti}.}:
\begin{eqnarray}
f(s^+)   \equiv  \mbox{Breit-Wigner}(D_0^{\ast +}\to D^+\pi^0) & = & \frac {m_{D_0^\ast}\sqrt{8\pi\BR_0\frac {\Gamma_0}{p^\ast_0}}}{s^+-m_{D_0^\ast}^2+ i m_{D_0^\ast} \Gamma_0} \label{bw}\\
A(\bar  B^0 \to D^+  D^{\ast -}_0 \to D^+ D^- \pi^0 ) & = & 
T_1\,\, f(s^-) \label{amp1}\\
A(\bar  B^0 \to  D^{\ast +}_0 D^- \to D^+ D^- \pi^0 ) & = & 
T_2\,\, f(s^+) \\ 
A(B^0 \to   D^- D^{\ast +}_0 \to D^+  D^- \pi^0 ) & = & 
T_1\,\, f(s^+) \\
A(B^0 \to D^{\ast -}_0 D^+ \to D^+ D^- \pi^0 ) & = & 
T_2\,\, f(s^-) \label{amp4}
\end{eqnarray}
where $\BR_0$ is the branching ratio $D^{\ast +}_0\to D^+\pi^0$, $ p^\ast_0$ is the momentum of the $D$ in the $D_0^\ast$ rest frame,
and where we have used the \CP-identities in Wolfenstein phase convention
\footnote{Any constant phase in the normalization of the Breit-Wigner functions can be incorporated in the definition of $T_1$ and $T_2$.}:
\begin{eqnarray}
M(\bar  B^0 \to D^+  D^{\ast -}_0) & = & M( B^0 \to D^-  D^{\ast +}_0) \equiv T_1 \label{t1}\\
M(\bar  B^0 \to D^{\ast +}_0 D^-) & = & M( B^0 \to D^{\ast -}_0  D^+)\equiv T_2. \label{t2}
\end{eqnarray}
Our convention is to write first, in the final state, the meson which is directly coupled to the initial meson via a current form factor, and second, the one which is emitted by the $W$ meson. The numerators in~(\ref{bw}) have been fixed by imposing that in the narrow width limit the integrated three-body decay width for $\bar  B^0 \to D^+  D^{\ast -}_0 \to D^+ D^- \pi^0 $ coincides with the two-body one $\bar  B^0 \to D^+  D^{\ast -}_0$ times $\BR_0$. For resonances of spin $J$ there is an additional multiplicative factor of $Y_J^0(\theta^\ast,0)/\sqrt{4\pi}$, where $\theta^\ast, \varphi^\ast$ are the spherical angles of the decay products in the resonance rest frame. Angular momentum conservation selects $m=0$ and hence there is no dependence on $\varphi^\ast$.

Defining 
\begin{eqnarray}
{\rm G}_{\rm tot}&\equiv&|T_1|^2+|T_2|^2 \label{gamma1}\\
R&\equiv&\frac{1}{{\rm G}_{\rm tot}}\left ( |T_1|^2-|T_2|^2 \right ) \label{defR}\\
D&\equiv&\sqrt{1-R^2}=\frac{2}{{\rm G}_{\rm tot}}|T_1||T_2|\label{defD} \\
\delta&\equiv&\mbox{Arg}\left ( T_1T_2^\ast \right )
\end{eqnarray}
we obtain:
\begin{eqnarray}
{\rm G}_0(s^+,s^-) & = & {\rm G}_{\rm tot}\left \{ |f(s^+)|^2 +|f(s^-)|^2 +2D\,\cos\delta\,
		\Re \left ( f(s^+)f(s^-)^\ast \right ) \right \} \label{0}\\
{\rm G}_{\rm c}(s^+,s^-) & = & {\rm G}_{\rm tot}\left \{ R\left [ |f(s^+)|^2 -|f(s^-)|^2 \right ]
 	-2D\,\sin\delta\,\Im \left ( f(s^+)f(s^-)^\ast \right ) \right \} \label{cos}\\
{\rm G}_{\rm s}(s^+,s^-) & = & {\rm G}_{\rm tot}\left \{-D\left [ \sin(2\beta+\delta)|f(s^+)|^2
+\sin(2\beta-\delta)|f(s^-)|^2 \right ] \right. \nonumber \\
             & - & 2\sin 2\beta\,\Re \left ( f(s^+)f(s^-)^\ast \right )
               -   \left. 2R\,\cos 2\beta\,\Im \left ( f(s^+)f(s^-)^\ast \right )  \right \}\label{gmix2} \end{eqnarray}

The important point is that by measuring the kinematical and time-\-de\-pen\-dence of the three-body decay, one extracts from~(\ref{0}--\ref{gmix2}) all the two-body complex amplitudes~(\ref{t1}--\ref{t2}), as well as $2\beta$, without discrete ambiguities in the 
general case. This is true whatever the number of contributing resonances is, as long as we are able to deal with the multi-parameters fit
\footnote{This statement is valid assuming the resonances are well known (mass, width, spin...). This could be a difficult experimental task, particularly for the $D^*_{0}$.}.

The amplitudes $T_1$ and $T_2$ are not in general related. From~(\ref{defR}), $R$ is thus typically of order 1.  Then one is lead to use the last term of the rhs of~(\ref{gmix2})
to measure $\cos 2 \beta$. If by some symmetry or by chance $R$ is small, $D$ is of order 1 (in fact both $R$ and $D$ are often of order 1). The first two terms in the rhs of~(\ref{gmix2}) may then be used. There the  $\cos 2 \beta$ term comes also from an interference between strong phases, but instead of the Breit-Wigner phases, it uses the strong phase $\delta$ of the decay $B\to D^{\ast\ast} \bar D$. This phase, which vanishes in the factorization approximation, might be small $\simeq 1/N_c$. To get it one has to use eqs.~(\ref{0}) and~(\ref{cos}) where it is measured thanks to the Breit-Wigner interference. This shows that the determination of $\cos 2\beta$ is overconstrained.

To summarize, the extraction of both the sine and cosine of the CKM phase rely on the important hypothesis (hopefully very reasonable) stated as follows: 
\begin{center}
{\it Over the Dalitz plot, no other \CP-even phases than \`a la Breit-Wigner,\\
or constants, are allowed.}
\end{center}

 In the $B\to D^+D^- \pi^0$ channel, $R$ is  indeed 
 presumably close to 1, as will be discussed in a model in Section~\ref{model}, because the 
 $B\to D$ transition is strongly favored over $B\to D^\ast_0$, while the current
  matrix elements to the vacuum are presumably of the same order of magnitude. On the contrary, for the $D_2^\ast$ contribution, one has $R = -1$ exactly in the
 factorization approximation as the $D_2^\ast$ does not couple to the current.
 
 From~(\ref{gmix2}), one sees also that whatever the values of $R$ and $D$ are, a
 $\sin 2\beta$ term is present. This term will not compete with the measurement of the $B\to J/\psi K_S$ asymmetry, but it can be used either as a cross-check or as an additional input which would help a multi-parameters likelihood fit.
 
 Now, the factor $f(s^+)f^\ast(s^-)$ will be studied in some detail in the coming subsection. It will be seen that the wanted effect grows like the resonance width and like its mass squared.
 This is why rather larger and heavier resonances are better. They should however not be too large in order
  to emerge from the background noise. {\it A priori}, the non-yet observed $D^\ast_0$ resonance is large enough $\sim $ 150 MeV, while the known $D_2^\ast$ is rather narrow
  $\sim$ 20 MeV.
   
   In this section we have neglected the penguin diagrams, although this is not
 really legitimate. Unhappily, the Dalitz plot analysis does not allow to
 separate penguin contributions from the tree ones, as it can be done in the $3\pi$ case~\cite{quinn},
 because for the $b \to c \bar c d$ transition , trees and and penguins 
 have the same $\Delta I=1/2$ isospin structure~\cite{sanda}.
  Of course it might be argued that since the aim is to 
 resolve a discrete ambiguity on $\beta$, a small penguin induced error is
  not a problem. But a penguin free channel would be safer, and this is 
  the theme of the next sections.  Before that let us in the next subsection work out a crude estimator of the number of $\cos 2 \beta$ sensitive events in the Dalitz plot.

\subsection{Effective Branching Ratio} \label{BReff}

 Throughout this letter we are interested in Breit-Wigner interferences, either
 to generate a $\cos 2\beta$ term, as in $D\bar D\pi$, $D\bar DK$ and $D\pi\pi$ final states
 (see~(\ref{gmix2})),
 or to produce $2\beta+\gamma$ dependence, as in $D\pi K$ final state where all the CKM angle dependence
 is proportional to the Breit-Wigner interferences (see~(\ref{2b+g})).
 
\begin{figure}
\centering
\includegraphics[width=.4\textwidth]{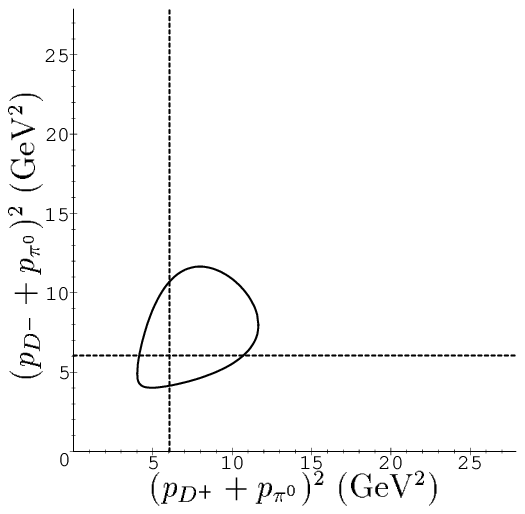}
\includegraphics[width=.4\textwidth]{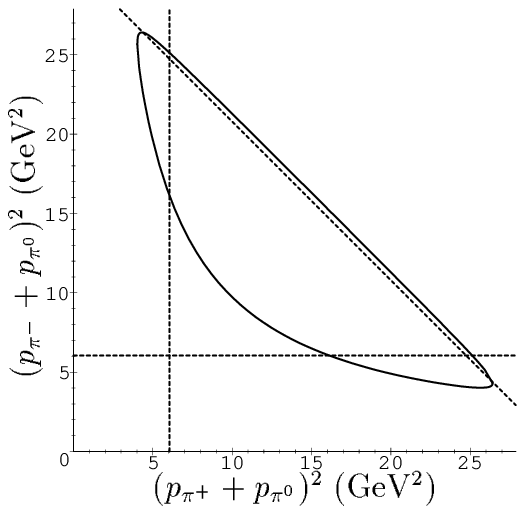}
\caption{\it To the left, the shape of the $B\to D\bar D\pi$ phase space (the straight lines represent the position of the $D^{\ast\ast}$ resonances). To the
right, the same for $B\to D\pi\pi$ (the third straight line represents the $\rho$). It is seen that the $D^{\ast\ast}/\bar D^{\ast\ast}$ resonances cross inside the $D\bar D\pi$ physical domain, and outside the $D\pi\pi$ one.}
\label{dalitzShape}
\end{figure}

Let us call $f_i$ the Breit-Wigner of the resonance $R_i$ the mass, width and spin of which is respectively $m_i$, $\Gamma_i$, $J_i$. Recall eq.~(\ref{gmix2}): the interference term $f_1 Y_1 f_2^\ast Y_2^\ast$ describes the fluctuations of the number of events which are due to $\cos 2\beta$. What is the size of this term ? $Y_i \equiv \sqrt{4\pi} Y_{J_i}(\theta_i^\ast,0)$ with $\theta_i^\ast$ 
 in the  rest frame of
 resonance $R_i$ corresponds to the angle between the resonance decay product momentum and the momentum of the $B$. When the resonance is in the $s^+$ channel, $\cos \theta^\ast_i$ depends linearly on $s^-$, and vice-versa.
 In practice the resonance crossing region corresponds to  $\theta^\ast_i\simeq 0$ which tends to enhance the interference effect for large spin of the resonances. To get a flavor of the result, let us first consider the simpler case
 of $J_1=J_2=0$:
 \begin{equation}
 N_{12}= \int ds^+ ds^- |\Im(f_1(s^+)f_2^\ast(s^-))| = 8 \pi m_1 m_2 \sqrt{\BR_1\BR_2\frac{\Gamma_1\Gamma_2}{p^\ast_1p^\ast_2}}\times I_{12}\label{num}
 \end{equation}
 where
 \begin{equation} I_{12}=\int dx dy \frac {|y-x|}{(x^2+1)(y^2+1)},\label{I}
 \end{equation}   
$\BR_i$ is the branching ratio of $R_i$ in the considered decay products, and where $p^\ast_i$ is
the final momentum of  these decay products in the resonance rest frame 
($p^\ast_i$ ranges from 300 to 500 MeV in the considered cases). We end up with a term proportional to $ 8 \pi m_1 m_2
\sqrt{\Gamma_1\Gamma_2/(p^\ast_1 p^\ast_2)}$ times a dimensionless integral.
The latter depends on the masses and widths through the integration bounds
which are typically of the form $(s_{max}-m_i^2)/(\Gamma_i m_i)$. A rapid inspection shows that the dimensionless integrals cannot depend on the phase space parameters and  widths more than logarithmically. So the main dependence is included in the dimensionfull factor: 
\begin{equation}F_{12}=8 \pi m_1 m_2 \sqrt{\BR_1\BR_2\frac{\Gamma_1\Gamma_2}{p^\ast_1p^\ast_2}}.
\label{F}
\end{equation}
 We immediately see that heavy resonances will be favored. This can be partially understood as the width of the resonance band in the Dalitz plot is $m\Gamma$ and not $\Gamma$ simply. The results of exact numerical integration of~(\ref{num}) will
be given for each channel in Tables~\ref{tableddpi},~\ref{tableddk},~\ref{tabledpipi},~\ref{tabledkpi}. It happens that the dimensionless integral is larger when 
the resonance crossing is inside the physical domain than when it is not. In Figure~\ref{dalitzShape}, we have plotted the shape of two representative phase spaces, namely $B\to D\bar D\pi$ and $B\to D\pi\pi$. In the former case the $D^{\ast\ast}/\bar D^{\ast\ast}$ cross well inside the physical domain, and $I_{12}$ is typically $\sim {\rm C}^{\rm st}\times\log\left[(s_{max}-m_i^2)/(m_i \Gamma_i)\right] \sim 5-20$, while in the latter the $D^{\ast\ast}/\bar D^{\ast\ast}$ cross well outside the physical domain, and $I_{12}$ is of order 1. 
This is expected since we are looking for an interference effect between these two resonances, but it is noticeable that even in the latter case, there is a sizeable dependence on 
$\cos 2 \beta$ located far from the resonance interference region. This feature, already seen in the three-pion case~\cite{quinn,babarbook}, is due to the fact that  the Breit-Wigner formulae, when they are not squared, as in~(\ref{num}), decrease slowly away from the resonance center. However, as already discussed, the domains too far away from the resonance regions need much care since other resonances, not considered in the fit, may be dominant in the latter regions. When the resonances do cross
in the physical domain, a very conservative estimate of the $\cos 2\beta$ sensitive events is equivalent to take $I_{12}=1$ in~(\ref{num}), or 
\begin{equation}
I_{\rm 12,cons}\simeq 4\pi \sqrt{Y^0_{J_1}(0,0)\,Y^0_{J_2}(0,0)}=\sqrt{(2J_1+1)(2J_2+1)}\label{Icons}
\end{equation} in the case of resonances of spins $J_1$, $J_2$. When they do not cross, the conservative estimate is simply 0.

Now we are able to define an Effective Branching Ratio ($\BR_{\rm eff}$) for the
contribution of $D_0^\ast$ to the $\cos 2\beta$ term (see eqs.~(\ref{gmix2}) and~(\ref{num})):  
\beq
\BR_{\rm eff}=\left[\BR(\bar B^0\to D_0^{\ast +}D^-)+\BR(\bar B^0\to D^+D_0^{\ast -}\right]
\times R \times \frac{N_{12}}{32\pi^2m_B<p>} ,
\eeq
with the two-body $\BR$ calculated in the zero width limit (in the following, we use 
the factorization assumption to compute these \BR) and
\beq
<p>=\frac{p_1|T_1|^2+p_2|T_2|^2}{|T_1|^2+|T_2|^2}.
\eeq
$p_1$ (resp. $p_2$) is the momentum in the rest frame of the $B$ of the resonance 1
(resp. resonance 2). 

 The fact that oscillations generate the $\cos 2 \beta$ factor is obvious from~(\ref{osc})
since ${\rm G}_{\rm s}$ multiplies $\sin \Delta m\,t$. We must expect an additional suppression due to these oscillations. However, to get an idea of this suppression one integrates over time $\exp(-\Gamma t)\sin \Delta m\,t$, which gives $x/(1+x^2)\simeq 0.48$ where $x\equiv \Delta m/\Gamma\simeq 0.73$. We will neglect this factor 1/2 in view of the crudeness of our estimate.

For the $B\to D^+D^-\pi^0$ Dalitz plot, the Effective Branching Ratios concerning the
contributions of $D_0^\ast$ and $D_2^\ast$ are shown in Table~\ref{tableddpi}. The
lower bound corresponds to the conservative estimate for $I_{12}$ (see~(\ref{Icons}))
while the upper bound corresponds to the full numerical integration of~(\ref{I}). The
two-body non-leptonic branching ratios have been calculated in the factorization assumption,
as explained in section~\ref{facto}.
The last line of the table sums all the contributions taking into account the detection
efficiency~\cite{MH} of the particles in the final state.
\begin{table}
\begin{center}
\begin{tabular}{|c|c|c|c|} \hline
\multicolumn{4}{|c|}{$B^0(t)\to D^+D^-\pi^0$\hspace{1cm} $\Phi=2\beta$} \\ \hline\hline
Crossing & $\BR_{\rm eff}$ for $\cos \Phi$ & Total \BR & Detection Efficiency \\
$D_0^{\ast +}/D_0^{\ast -}$ & (0.2--3)$\times 10^{-5}$ & 21$\times 10^{-5}$ & 0.005--0.01 \\ \cline{4-4}
$D_2^{\ast +}/D_2^{\ast -}$ & (0.01--0.04)$\times 10^{-5}$ & 1.9$\times 10^{-5}$ & \multicolumn{1}{c}{} \\
\cline{1-3}
Total & (0.1--3)$\times 10^{-7}$ & \multicolumn{2}{c}{} \\
\cline{1-2}
\end{tabular}
\caption{\it Effective Branching Ratio for $\cos2\beta$ in the $B\to D^+D^-\pi^0$ channel.}
\label{tableddpi}
\end{center}
\end{table}

For comparison, the Effective Branching Ratio for the measurement of $\cos2\alpha$ in the $3\pi$ case is about 10$^{-6}$. In this case the $\rho^+/\rho^-$ crossing lies just at the border of the physical domain. From Table~\ref{tableddpi}, we see that the possibility of measuring $\cos 2\beta$ in $B\to D^+D^-\pi^0$ channel relies strongly on the contribution of the $D_0^\ast$ (because the transition $\bar B^0\to D^+D_0^{\ast -}$ dominates over $\bar B^0\to D^+D_2^{\ast -}$ and $\bar B^0\to D_2^{\ast +}D^-)$ and thus on our ability to describe accurately this resonance. In conclusion, this channel seems to be not accessible to an $e^+e^-$ $B$-factory and the presence of a $\pi^0$ is problematic for an hadronic machine. However, it is straightforward to generalize this idea to $B\to D^{(\ast)+}D^{(\ast)-}\pi^0$ channels, for which the detection efficiency will be much more favorable.

In the following, we will consider cases where the two crossing resonances are not
necessarily \CP-conjugate. The calculation of $\BR_{\rm eff}$ follows the same line and only the numerical results will be presented in the next sections.

\section{ $B\to D^+D^-K_S$} \label{ddk}
 
At the quark level, the transition generating this channel is $b\to c\bar cs$, that is, {\it the same as in $B\to J/\Psi K_S$}. This has two important consequences: First, it is a class-I, Cabibbo-favored channel (proportional to $|a_1V_{cb}V_{cs}|$). Second, the penguin pollution is suppressed
\footnote{We have used CKM unitarity to incorporate the $|V_{tb}V_{ts}|$ contribution in the "tree" part proportional to $|V_{cb}V_{cs}|$ and in the "penguin" part proportional to $|V_{ub}V_{us}|$.}
 by $ |V_{ub}V_{us}|/|V_{cb}V_{cs}| \sim 2\times 10^{-2} $. The kinematics of this decay is very similar to $B\to D^+D^-\pi^0$, replacing the $D^{\ast\ast}$ resonances by the $D_s^{\ast\ast}$.  Due to its flavour structure, the $D_s^{\ast\ast}$ contribute only when connected to the $W$ propagator. Thus the contribution of the $D_{s2}^{\ast}$ is expected to be very small (it is zero in the factorization approximation), and the non-yet observed $D_{s0}^{\ast}$ should dominate the decay. We assume $m_{D_{s0}^\ast}=2.5$ GeV; for the width, we take $\Gamma_{s0}=150$ MeV and assume that $DK$ saturates the decay~\cite{kaidalov}. The equations~(\ref{0}--\ref{gmix2}) apply, with $ R=1,\ D=0 $. The calculation of the Effective Branching Ratio, summarized in Table~\ref{tableddk}, is very encouraging ($\sim 10^{-4}$), although it has once more to be corrected by the $D$ meson detection efficiency. In any case,
 this channel should be about 20 times more favorable than the $B\to D^+D^-\pi^0$ one.

Finally, this analysis is not expected to be generalizable to the analogous decay with one (or two) $D_s^\ast$ in the final state, as the $D_{s0}^{\ast}$ does not decay in $D_s^\ast K$, the $D_{s1}^{\ast}$ is expected to be below the $D_s^\ast K$ threshold, the $D_{s1}$ is too narrow ( $\Gamma < 2.3$ MeV), and the branching ratio for the $D_{s2}^{\ast}$ is expected to be suppressed.
\begin{table}
\begin{center}
\begin{tabular}{|c|c|c|c|} \hline
\multicolumn{4}{|c|}{$B^0(t)\to D^+D^-K_S$\hspace{1cm} $\Phi=2\beta$} \\ \hline\hline
Crossing & $\BR_{\rm eff}$ for $\cos \Phi$ & Total \BR & Detection Efficiency \\
$D_{s0}^{\ast +}/D_{s0}^{\ast -}$ & (0.6--6)$\times 10^{-4}$ & 35$\times 10^{-4}$ & 0.005--0.01 \\ \hline
Total & (0.3--6)$\times 10^{-6}$ & \multicolumn{2}{c}{} \\
\cline{1-2}
\end{tabular} 
\caption{\it Effective Branching Ratio for $\cos 2\beta$ in the $B\to D^+D^-K_S$ channel.}
\label{tableddk}
\end{center}
\end{table}

\section{ $B\to D^0_{CP} \pi^+\pi^-$} \label{dpipi}

By $D^0_{CP}$ we mean a $D^0$ meson which eventually decays into a \CP-eigenstate,
for example $K_S \pi^0$,  $\pi^+\pi^-$, $\pi^0\pi^0$, $K^+ K^-$ $K^0 \bar  K^0$, i.e. a few \% of the $D^0$ are used. However, compared to the situation in the preceding sections, this loss in statistics is similar to the two $D$ mesons detection efficiency. Furthermore, as $B\to D^+D^-K_S$, $B\to D^0_{CP} \pi^+\pi^-$ is Cabibbo-favored (proportional to $|V_{cb}V_{ud}|$). The {\it very noticeable advantage of $B\to D^0_{CP} \pi^+\pi^-$ is that there 
is no penguin contribution at all}. Indeed the weak decay at the quark level is 
of the type $b \to c \bar u d$ and obviously no penguin operator can contribute to such a transition with four changes in flavor.

The analysis proceeds then in a way parallel to the one in the preceding sections.
The two-body channels will be:
\begin{equation}
\bar  B^0 \to D^{\ast\ast +} \pi^-\quad\hbox{and}\quad \bar  B^0 \to \rho^0D^{0}, 
\end{equation}
and their \CP-conjugates.
The channels $\bar B^0 \to \pi^+ D^{\ast\ast -} $ and $\bar B^0 \to \rho^0 \bar D^0$ are doubly-Cabibbo-suppressed
\footnote{Proportional to $|V_{ub}V_{cd}|$, they would introduce a very small dependence on the angle $2\beta+\gamma$, and could in principle be incorporated in the analysis. However, the $B\to D^\pm\pi^\mp K_S$ channel is much more favorable in this respect.} and we will forget them. Hence the ratio $R$~(\ref{defR})
relevant for the crossing between $\bar B^0 \to D^{\ast\ast +} \pi^-$ and $B^0\to  D^{\ast\ast -} \pi^+$ is equal to 1. The branching ratios recently measured by
CLEO~\cite{cleo1} for $B\to D_1 \pi$ and $B\to D_2^\ast \pi$ are of order 10$^{-3}$.

It should be noted here that the $D^{\ast\ast}/\bar D^{\ast\ast}$ crossings are
outside the physical domain. We then better consider the
$D^{\ast\ast}/\rho$ crossings which are inside the physical domain. The problem now
is that it is difficult to estimate accurately the ratio:
\begin{equation}
 \frac {|A(\bar B^0 \to  \rho^0D^{0})|}
 {|A(\bar B^0\to D^{\ast\ast +}\pi^-)|} . \label{rata}
\end{equation}
$A(\bar B^0\to D^{\ast\ast +}\pi^-)$ is suppressed by the $\tau_{1/2}$ or $\tau_{3/2}$ form factors,
while $A(\bar B^0 \to \rho^0D^{0})$ is color-suppressed. The ratio~(\ref{rata}) is
 expected to be of order 1, as happens in the factorization assumption, with the form factors computed in our model 
described in section~\ref{facto}.
\begin{table}
\begin{center}
\begin{tabular}{|c|c|c|c|} \hline
\multicolumn{4}{|c|}{$B^0(t)\to D^0_{\CP}\pi^+\pi^-$\hspace{1cm} $\Phi=2\beta$} \\ \hline\hline
Crossing & $\BR_{\rm eff}$ for $\cos \Phi$ & Total \BR & Detection Efficiency \\
$D_0^{\ast +}/D_0^{\ast -}$ & (0.1--0.4)$\times 10^{-5}$ & 16$\times 10^{-5}$ & 0.005--0.01 \\ \cline{4-4}
$D_2^{\ast +}/D_2^{\ast -}$ & $\sim$ 0.4$\times 10^{-5}$ & 120$\times 10^{-5}$ & \multicolumn{1}{c}{} \\
$D_0^{\ast +}/\rho^0$ & (0.1--2)$\times 10^{-5}$ & 32$\times 10^{-5}$ & \multicolumn{1}{c}{} \\ 
$D_2^{\ast +}/\rho^0$ & (0.2--4)$\times 10^{-5}$ & 130$\times 10^{-5}$ & \multicolumn{1}{c}{} \\ \cline{1-3}
Total & (0.05--1)$\times 10^{-6}$ & \multicolumn{2}{c}{} \\
\cline{1-2}
\end{tabular}
\caption{\it Effective Branching Ratio for $\cos 2\beta$ in the $B\to D^0_{\CP}\pi^+\pi^-$ channel.}
\label{tabledpipi}
\end{center}
\end{table}

The Table~\ref{tabledpipi} shows the Effective Branching Ratio for this channel, which is comparable to the one corresponding to the measurement of $\cos2\alpha$ through the $3\pi$ channel. In spite of the
relative narrowness of the $D_2^\ast$, the $D_2^\ast/\rho$ crossing is predicted to be one dominant contribution to the
measurement of $\cos2\beta$, thanks to the spin effects, the width of the $\rho$ and the large
branching ratio for $B\to D_2^\ast\pi$ compared to $B\to D_0^\ast\pi$. This is nice, because the $D_2^\ast$, which is already seen in semi-leptonic and non-leptonic $B$ decays, will be 
well known in a close future.

\section{$B\to D^{\pm} \pi^{\mp} K_S$} \label{dkpi}

In the case of the $D^+\pi^-K_S$ final state, the intermediate resonant channels are
\beq
\bar B^0 \to D^+ K^{{\ast}^-} \not\!\from  B^0 \ ,\ \ \ \bar B^0 \not\to \pi^-D^{\ast\ast +}_s \from B^0 \ ,\ \ \ \bar B^0 \to K_SD^{\ast\ast 0} \from B^0.
\eeq
The $K^{\ast}$ contributes only to diagrams proportional to $|V_{us}V_{cb}|$ ($={\cal O}(\lambda^3)$ in Wolfenstein parametrization)
while the $D^{\ast\ast}_s$ contribute only to diagrams proportional to
$|V_{cs}V_{ub}|$ ($={\cal O}(\lambda^3)$ too). Since, at the quark level, the weak decays are $b\to c s \bar u$ and $b \to u s
\bar c$, {\it there is no penguin contribution in this
channel} with four flavor changes. This very
interesting channel measures $\cos(2\beta+\gamma)$ and $\sin(2\beta+\gamma)$. Indeed, in Wolfenstein
parametrization, the CKM matrix element $V_{ub}$ contains the phase $\gamma$ while the
$B^0-\bar B^0$ mixing is proportional to $\exp(-2 i \beta)$. To our knowledge, the only other ways to measure $\sin(2\beta+\gamma)$ is to look at the 
very small \CP-asymmetry in the dominant channel $B\to D(\bar D)\pi$, the larger \CP-asymmetry in the color-suppressed channel
$B\to K_SD^0(\bar D^0)$, or in some semi-inclusive $b\to u\bar cs+c\bar us$ decays~\cite{dunietz}; no other method to measure directly $\cos(2\beta+\gamma)$ has been proposed in the literature.

As this channel is a little different from the others presented in the previous sections, we 
rewrite equations~(\ref{0}--\ref{gmix2}) for the $D^+\pi^- K_S$ final state. Neglecting for simplicity the neutral mode $K_SD^{\ast\ast 0}$, the following channels contribute:
\begin{eqnarray}
A_{D_s^{\ast\ast}}= A(B^0\to \pi^-D^{{\ast +}}_{s0})
&\equiv&T_{D_{s0}^\ast} e^{i\gamma} \\
\bar{A}_{K^\ast}= A(\bar{B}^0\to D^+K^{{\ast}^-})
&\equiv& T_{K^\ast}
\end{eqnarray} 
where we made use of the fact that the tensor (J=2) $D_{s2}^\ast$ particle 
cannot couple to currents from the vacuum, and thus $A(B^0\to \pi^-D^{{\ast +}}_{s2})$ is suppressed. Defining:
\begin{eqnarray}
{\rm G}_{\rm tot}&\equiv&|T_{D_{s0}^\ast}|^2+|T_{K^\ast}|^2 ,\\
R&\equiv&\frac{1}{{\rm G}_{\rm tot}}\left ( |T_{D_{s0}^\ast}|^2-|T_{K^\ast}|^2 \right ) ,\\
D&\equiv&\sqrt{1-R^2}=\frac{2}{{\rm G}_{\rm tot}}|T_{D_{s0}^\ast}||T_{K^\ast}| ,\\
\delta&\equiv&\mbox{Arg}\left ( T_{D_{s0}^\ast}T_{K^\ast}^\ast \right ) ,\\
g_+&\equiv&\mbox{Breit-Wigner}(D^{{\ast +}}_{s0}\to D^+K_S) ,\\
h_-&\equiv&\mbox{Breit-Wigner}(K^{\ast^-}\to K_S\pi^-),
\end{eqnarray} 
we get
\begin{eqnarray}
{\rm G}_0 & = & {\rm G}_{\rm tot} \left [ \frac{1+R}{2} |g_+|^2 
+ \frac{1-R}{2} |h_-|^2 \right ] ,\\
{\rm G}_{\rm c} & = & {\rm G}_{\rm tot} \left [ \frac{1+R}{2} |g_+|^2 
- \frac{1-R}{2} |h_-|^2 \right ] ,\\
{\rm G}_{\rm s} & = & {\rm G}_{\rm tot}D \left [-\cos (\delta+2\beta+\gamma)
		\Im \left ( g_+ h^\ast_- \right )
				    -\sin (\delta+2\beta+\gamma)
		\Re \left ( g_+ h^\ast_- \right ) \right ]. \label{2b+g}
\end{eqnarray} 

The $B^0(t)\to D^+\pi^-K_S$ time-dependent decay allows a measurement
of the angle $\delta+2\beta+\gamma$, while $\bar{B}^0(t)\to D^-\pi^+K_S$ leads
to $\delta-2\beta-\gamma$. In the end, one obtains {\it separately} the \CP-violating phase $2\beta+\gamma\equiv \pi+\beta-\alpha$
and the \CP-conserving phase $\delta$. Once $2\beta$ is known (cf. above), this could
be a very interesting measurement of $\gamma$.

The contribution of the $K^\ast(892)$ resonance to the $2\beta+\gamma$ terms suffers from its
relatively small mass and width and more importantly from the fact that the $K^\ast(892)/D_{s0}^\ast$ crossing is outside the physical domain. Thus it is important to take into account
higher $K^\ast$ resonances such as the broad (287 MeV) scalar $K_0^\ast(1430)$ (the contribution of the $K_2^\ast$, although interesting, vanishes in the factorization assumption):
these crossings lie just at the border of the physical domain. From Table~\ref{tabledkpi}, the measurement of both $\sin(2\beta+\gamma)$ and $\cos(2\beta+\gamma)$ from $D\pi K$ analysis 
should be a task comparable to the measurement of $\cos2\alpha$ from $3\pi$ channel, although 
some care has to be taken deciding which resonances should be cut or retained.
\begin{table}
\begin{center}
\begin{tabular}{|c|c|c|c|c|} \hline
\multicolumn{5}{|c|}{$B^0(t)\to D^{\pm}\pi^{\mp}K_S$\hspace{1cm} $\Phi=2\beta+\gamma$} \\ \hline\hline
Crossing & $\BR_{\rm eff}$ for $\cos \Phi$ & $\BR_{\rm eff}$ for $\sin \Phi$ & Total \BR & Detection Efficiency \\
$D_{s0}^{\ast +}/K^{\ast -}$ & (0.04--0.2)$\times 10^{-5}$ & (0.04--0.4)$\times 10^{-5}$ & 27 $10^{-5}$ & 0.05--0.1 \\ \cline{5-5}
$D_{s0}^{\ast +}/K_0^{\ast -}$ & (0.1--2)$\times 10^{-5}$ & (0.1--2)$\times 10^{-5}$ & 27 $10^{-5}$ & \multicolumn{1}{c}{} \\ \cline{1-4}
Total & 0.07--2 $10^{-6}$ & 0.07--3 $10^{-6}$ & \multicolumn{2}{c}{} \\
\cline{1-3}
\end{tabular}
\caption{\it Effective Branching Ratio for $\cos(2\beta+\gamma)$ and  $\sin(2\beta+\gamma)$ in the $B\to D^{\pm}\pi^{\mp}K_S$ channel.}
\label{tabledkpi}
\end{center}
\end{table}

\section{Another $\cos2\beta$ channel: $B\to K_S K_S K_L$} \label{kkk}

Finally we present as a speculative idea the $B\to K_S K_S K_L$ Dalitz plot
where the resonant channels to be considered could be: $B\to \phi K_S$ and 
$B\to f_0(980) K_L$. From Bose statistics  $\phi$ decays into $K_L K_S$ in P-wave,
while $f_0(980)$ decays into $K_S K_S$ or as well $K_L K_L$  in S-wave. {\it Only penguin diagrams} contribute to these decays, which are real in Wolfenstein parametrization, except for $\lambda^2$ suppressed long-distance penguins, and thus the mixing angle $2\beta$ is measured by the Dalitz plot. 
The narrowness of the $\phi$ resonance may look worrying for this to work, the nature of the $f_0(980)$ is unclear ($q \bar q$ or ``molecule''), and anyhow  a closer scrutiny of this idea is needed. The crude estimator described in section~\ref{BReff} gives in this case an Effective Branching Ratio of order 10$^{-8}$ or less (assuming $B\to \phi K_S \sim B\to f_0 K_L \sim 10^{-5}$).

\section{A simple model for $B\to D^+ D^- \pi^0$} \label{facto}
\label{model}

First we must repeat that our aim in this letter is mainly illustrative. 
We will use the factorization hypothesis for the non-leptonic decays. We will assume Heavy Quark Symmetry for the wanted heavy-to-heavy form factors and leptonic decay amplitudes. Finally we take the form factors as computed recently in a class of relativistic quark models~\cite{vincent}. 
Let us first give a brief account of the form factors and parameters we have used
\footnote{In~(\ref{ffBT}), and contrary to ref.~\cite{vincent}, we include a factor 0.9 in $\xi(w)$ to take into account perturbative and non-perturbative corrections to $\xi(1)$, thus avoiding overestimates of the rates.}:
\begin{equation}
\xi(w)=0.9\left(\frac 2 {1+w}\right)^2,\ \tau_{1/2}(w)=0.23 
\left(\frac 2 {1+w}\right)^{1.7},\ \tau_{3/2}(w)=0.54
\left(\frac 2 {1+w}\right)^{3},
\label{ffBT}
\end{equation}
were we use the form factors in the notation of~\cite{isgur} and 
the indices 1/2 (3/2) refer to the angular momentum $j$ of the light quanta 
in the $D^{\ast\ast}$ mesons, which is conserved in the heavy mass limit.

For the leptonic decay constants we take
\begin{equation}
f_D=f_{D_{1/2}}=200\ {\rm MeV},\qquad f_{D_{3/2}} =0 .
\end{equation}
The vanishing of  $f_{D_{3/2}}$ is a general consequence of the heavy mass limit~\cite{decay}.

The non-leptonic decays $\bar B^0\to D^{\ast\ast +} D^-$ and 
$\bar B^0\to D^+D^{\ast\ast -}$ are computed via the factorization hypothesis:
\begin{eqnarray}
M(\bar  B^0 \to D^+   D^{\ast -}_0 ) & = & {\rm K}_{db}\,\,a_1f_{D}^{(1/2)} (m_B-m_D)\sqrt{m_D m_B}\,(w+1)\xi(w) ,\\
M(\bar  B^0 \to D^{\ast +}_0 D^-) & = & {\rm K}_{db}\,\,a_12 f_{D}
(m_B+m_{D_0^\ast})\sqrt{m_{D^\ast_0}\,m_B}\,(w-1)\tau_{1/2}(w) ,\\
M(\bar  B^0 \to D^{\ast +}_2 D^-) & = & {\rm K}_{db}\,\,a_1\sqrt{2}  f_{D}
(m_B+m_{D_2^\ast})\sqrt{m_{D^\ast_2}\,m_B}\,(w^2-1)\tau_{3/2}(w) ,\\
M(\bar B^0\to D^+D_2^{\ast -}) & = & 0 ,
\end{eqnarray}
where $ {\rm K}_{db}=\frac{G_F}{\sqrt{2}} V_{cd}^\ast V_{cb} $ and the last equality follows from the fact that the $D^\ast_2$ has no matrix element to the vacuum due to $f_{D_{3/2}}=0$~\cite{decay}. 

 The three-body amplitudes are obtained by multiplying the above amplitudes with the Breit-Wigner formula:
\begin{eqnarray}
A(\bar  B^0 \to D^+  D_0^{\ast -}\to D^+  D^- \pi^0 ) & = & 
M(\bar  B^0 \to D^+D^{\ast -}_0)\,\, \frac {m_{D_0^\ast}\sqrt{8\pi \BR_0\frac {\Gamma_0}{p^\ast_0}}}{s^--m_{D_0^\ast}^2+ i m_{D_0^\ast} \Gamma_0},\nn\\  \hbox{with} \quad w & = & \frac{m_B^2+m_D^2-s^-}{2 m_B m_D}\label{55}\\
A(\bar  B^0 \to  D_0^{\ast +}  D^-\to D^+   D^- \pi^0 ) & = & 
M(\bar  B^0 \to D_0^{\ast +}  D^-) \frac {m_{D_0^{\ast}}\sqrt{8\pi \BR_0\frac {\Gamma_0}{p^\ast_0}}}{s^+-m_{D_0^\ast}^2+ i m_{D_0^\ast} \Gamma_0},\nn \\  \hbox{with}\quad  w & = & \frac{m_B^2+s^+-m_D^2}{2 m_B m_{D_0^\ast}}\\
A(\bar  B^0 \to  D_2^{\ast +} D^-\to D^+  D^- \pi^0 ) & = & 
M(\bar  B^0 \to D_2^{\ast +}D^-) \frac {m_{D_2^\ast}\sqrt{8\pi \BR_2\frac {\Gamma_2}{p^\ast_2}}h(s^+,s^-)}{s^+-m_{D_2^\ast}^2+ i m_{D_2^\ast} \Gamma_2}, \nn\\  \hbox{with}\quad w & = & \frac{m_B^2+s^+-m_D^2}{2 m_B m_{D_2^\ast}}\label{60}
\end{eqnarray}
 where $p^\ast_0$ ($p^\ast_2$) is the momentum of the final mesons in the rest frame of the decaying $D^\ast_0$ ($D^\ast_2$).  $\BR_0$ is the branching ratio $D^{\ast +}_0\to D^+\pi^0$, i.e. just a Clebsch-Gordan coefficient if we assume that $D\pi$ saturates the $D^\ast_0$ decay: $\BR_0=1/3$. $\BR_2$ takes into account an additional ratio of 70\% of $(D^\ast_2\to D\pi)/(D^\ast_2 \to D\pi+D^\ast \pi)$:
 $\BR_2=0.23$ and  finally the function 
\begin{equation}
h(s^+,s^-) = \frac{\sqrt{5}}2 \left (3 \cos^2 \theta^\ast-1\right)
\end{equation}
is proportional to $Y^0_2(\theta^\ast,\phi^\ast=0)$ function describing the D-wave decay of the $D^\ast_2$, $\theta^\ast$ being the angle of the $D$ meson with respect to the $B$ momentum in the $D^\ast_2$ rest frame.
The Breit-Wigner functions are normalized, as indicated in section~\ref{ddpi}, to recover the two-body decay widths in the narrow width limit. Notice that in~(\ref{55}--\ref{60}), the two-body amplitudes do depend on $(s^+,s^-)$ via the dependence of the form factors on $w$. It is often assumed in the fits~\cite{quinn,babarbook,frabetti} that the two-body amplitudes are constant, and admittedly it is difficult to avoid such an assumption in the present status of our knowledge, but it must be asserted that in reality they are only approximately constant, as we find in our model, as would be imposed by analyticity, etc. This is a source of theoretical error in the extraction of the CKM angles: its estimation is beyond the scope of this letter.

\begin{figure}
\centering
\includegraphics[width=.4\textwidth]{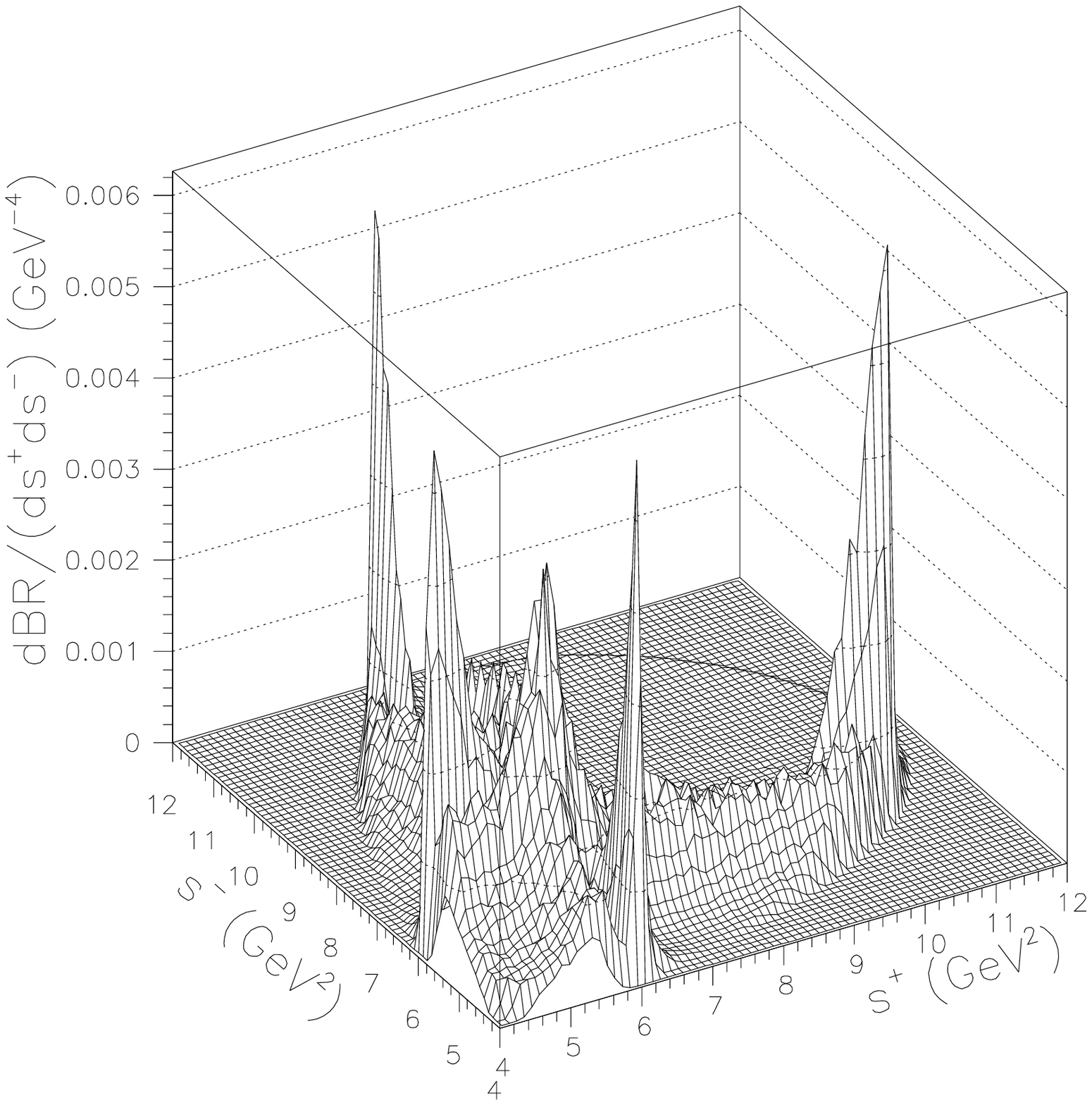}
\includegraphics[width=.4\textwidth]{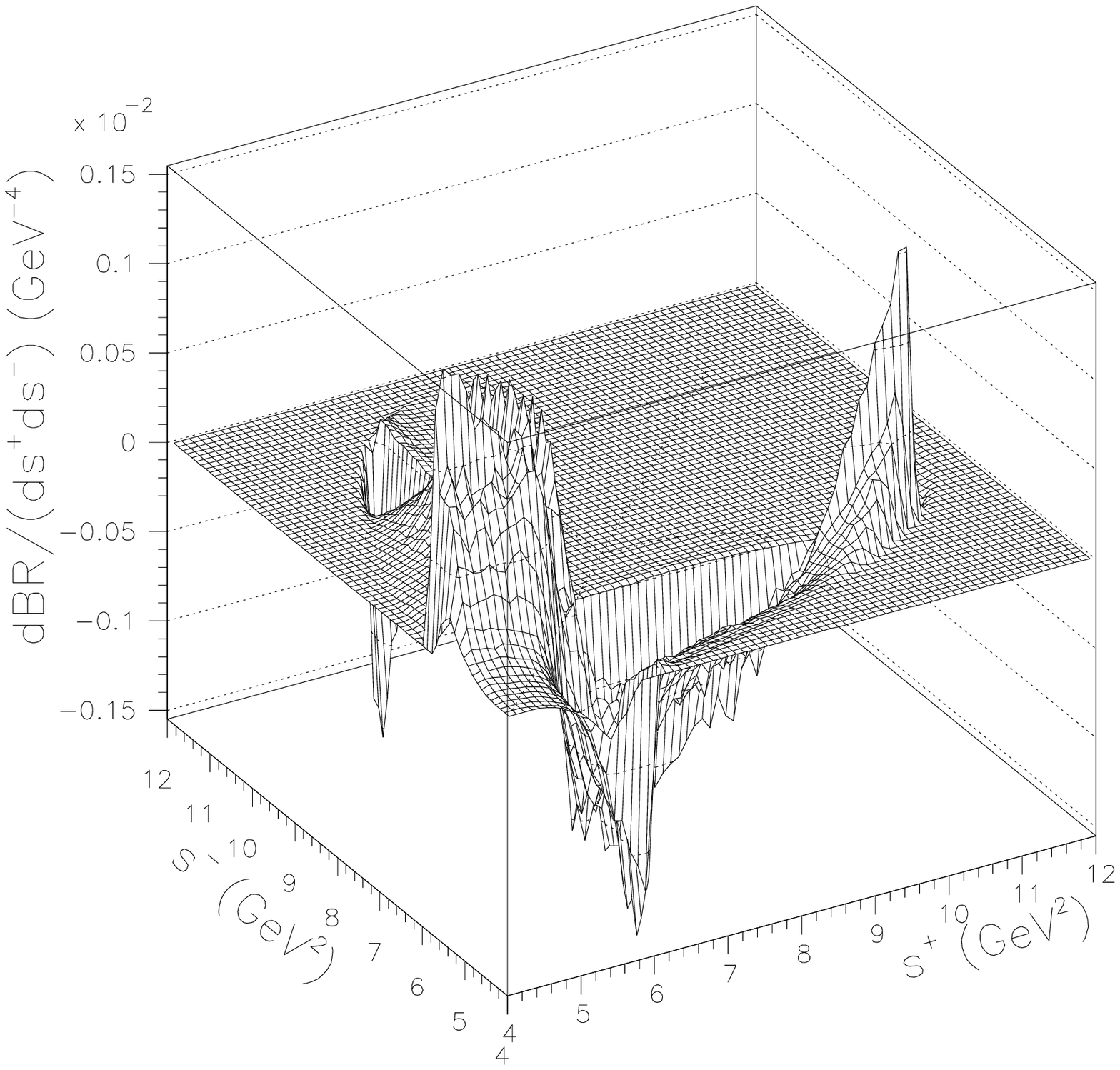}
\caption{\it $B^0(t)\to D^+D^-\pi^0$ from our simple
model. To the left, the differential branching ratio~(\ref{br}) for $\Delta m\,t=\pi/2, \beta=\pi/6$. To the right, also for $\Delta m\,t=\pi/2$ the differential branching ratio~(\ref{br}) for $\beta=\pi/6$ minus the one for $\beta=2\pi/6$. This difference isolates the $\cos 2\beta$ term.}
\label{dalitzPlot}
\end{figure}

Then
\beqa
{\cal A}(s^-,s^+) & = & \bar{\cal A}(s^+,s^-)  =  A(\bar  B^0 \to D^+  D_0^{\ast -}\to D^+  D^- \pi^0 )\\ \nn
& + & A(\bar  B^0 \to  D_0^{\ast +}  D^-\to D^+   D^- \pi^0 ) + A(\bar  B^0 \to  D_2^{\ast +} D^-\to D^+  D^- \pi^0 )
\eeqa
and the differential branching ratio is given by
\begin{equation}
d\,\BR(B^0(t)\to D^+ D^- \pi^0)=\frac 1 {8\pi^3}\frac {1}{32 m_B^3}\tau_{B}|A(s^+,s^-;t)|^2 ds^+ds^-\label{br}
\end{equation}
where $\tau_B= 1.56\ 10^{-12}$ s is the neutral $B$ meson life time and $|A(s^+,s^-;t)|^2$ is given by~(\ref{osc}).
The differential branching ratio $d\,\BR/ds^+ds^-$ as a function of the Dalitz point $(s^+,s^-)$ is plotted on Figure~\ref{dalitzPlot}. It is seen clearly that the $\cos2\beta$ term
produces a significant effect.

\section*{Conclusion}

We have proposed several three-body $B_d$ decay channels from which $\cos2\beta$ can be
extracted. When mediated by heavy resonances, the interference
effects should be more important than in the light systems.

The $B\to D^+D^-K_S$ mode is the most promising as it is Cabibbo-dominant and
almost penguin-free. For the same reasons $B\to D^0_{\CP}\pi^+\pi^-$ is very interesting. Although
not competitive in its naive version, the $B\to D^+D^-\pi^0$ channel should be
directly generalizable to $B\to D^{\ast +}D^{\ast -}\pi^0$ and thus would benefit from
a bigger detection efficiency, though it suffers from irreducible penguin uncertainties.

A new clean method to measure $2\beta+\gamma$ is described, which requires the analysis of the penguin-free $B\to D^{\pm}\pi^{\mp}K_S$ Dalitz plots.

All the above charmed three-body decays might be detected at an $e^+e^-$ $B$-factory.
The hadronic machines will be also a good place to look at the three most interesting channels, namely $B\to D^+D^-K_S$, $B\to D^0_{\CP}\pi^+\pi^-$ and $B\to D^{\pm}\pi^{\mp}K_S$.

Finally, we notice that the penguin-induced $B\to K_SK_SK_L$ mode could also measure $\cos2\beta$ although the branching ratio might be too small.

As shown by the huge uncertainty of our crude calculations, further work is needed, along the line of $3\pi$ studies, to demonstrate the feasibility of these analyses.

\section*{Acknowledgements}

We warmly acknowledge Mehdi Benkebil, Fran\c cois Le Diberder, Yuval Grossman, Anne-Marie Lutz, 
St\'ephane Plaszczynski, Helen R. Quinn, Marie-H\'el\`ene Schune, Sophie Versill\'e, Guy Wormser and other \textsc{BaBar} collaborators, with whom discussions on the issues of this letter were initiated and developed. We also thank Luc Bourhis for help.


\end{document}